# MAST-Upgrade Divertor Facility and Assessing Performance of Long-Legged Divertors

G. Fishpool[1*], J. Canik[2], G. Cunningham[1], J. Harrison[1], I. Katramados[1], A. Kirk[1], M. Kovari[1], H. Meyer[1], R. Scannell[1] and the MAST-Upgrade Team[1]

[1]*Euratom/CCFE Fusion Association, Culham Science Centre, Abingdon, OX14 3DB, UK.*
[2]*Oak Ridge National Laboratory, P.O. Box 2008, Oak Ridge, TN 37831, USA.*

## Abstract

A potentially important feature in a divertor design for a high-power tokamak is an extended and expanded divertor leg. The upgrade to MAST will allow a wide range of such divertor leg geometries to be produced, and hence will allow the roles of greatly increased connection length and flux expansion to be experimentally tested. This will include testing the potential of the Super-X configuration [1]. The design process for the upgrade has required analysis of producing and controlling the magnetic configurations, and has included consideration of the roles that divertor closure and increasing magnetic connection length will play.

# Introduction

A demonstration reactor tokamak (DEMO) divertor will require innovation and enhanced understanding and predictive capability in order to manage the plasma-wall interaction, especially the heat flux, over and above that which has lead to a divertor design for ITER. Recent work has considered the role that increased flux expansion and/or divertor leg extension might play in allowing higher-performance core-plasmas to be operated without excessive erosion and/or damage of the divertor target [1-4]. Experimental investigation of such divertor configurations is necessary to both improve fundamental understanding of divertors in general, and to assess potential benefits. The upgrade of MAST will allow this for extended- and expanded- leg divertor geometries [5-7], including the Super-X of Valanju et al [1].

Following a brief review of the overall concept of the MAST upgrade divertor geometry, this paper presents specific analysis in respect of key divertor attributes. The first analysis sets the context for realistic connection length, the second considers the geometry of the divertor and the gas sealing in respect of divertor closure, and the third considers how the divertor geometric properties can be controlled through the plasma pulse.

# Geometry of the MAST upgrade

The upgrade to MAST is designed to generate up/down symmetrical magnetic configurations, with a corresponding pair of large, outer divertors with a high degree of closure, and which allow significant flexibility in respect to the geometry of the divertor legs contained therein – example configurations of which are shown in Figure 1. The design is specifically for a well controlled double null configuration, making use of the previously demonstrated large ratio of order 10 of low-field-side to high-field-side power in this configuration in MAST [8]. Vertical position control will allow fine-scale variation of up/down magnetic symmetry, for varying the up/down balance of power and particle loads, and exploration of single null configurations, though the inner strike areas have not been specifically engineered for this mode of operation.

The large volumes of the divertor chambers, and the proximity of the nose (see Figure 1) to the X-point, mean that the closure of the divertor is achieved largely independently of the location of the strike point. This feature reduces the coupling between the conditions of the main plasma, and those of the divertor plasma, and therefore should assist in understanding the behaviour of the divertor itself.

With 8 pairs of divertor coils in the upgrade, and a relatively large divertor chamber, a wide range of configurations can be explored, with a useful degree of decoupling in the variation of flux expansion at the target, strike point location, and connection length (here defined as between outer mid-plane and divertor target, along a magnetic field line). "Conventional" divertor configurations deposit the majority of the power on a frustum of carbon tiles at a nominal major radius of ~0.8m, as shown in the black flux-surfaces in Figure 2. The two plots to the right of the flux plot show the effect across the first 2cm of mid-plane SOL of the configuration in terms of both the connection length and the expansion expressed as the ratio between target wetted area (ignoring toroidal asymmetries) and mid-plane (horizontal) area, i.e. the increase in spreading of power across the target due simply to the magnetic geometry of the SOL. A Super-X type configuration deposits the majority of the (attached) power at ~1.6m radius (blue surfaces in Figure 2). In between these extremes are geometries with a variety of expansions, both at the target and further away, yielding a variety of connection lengths and expansions, as exemplified in the red surfaces shown in Figure 2.

The way in which the connection length is distributed between mid-plane and target for field lines starting from 10mm radially outward from the separatrix is shown in the far right plot of Figure 2 for the three configurations. Between the mid-plane and the height of the X-point there is about 5m of connection length for this field line, with a similar amount between X-point and target for the conventional case. All three cases are very similar above and around the X-point, and the examples show how the connection length between X-point and target can be increased several times over the value for the conventional equilibrium.

**Connection Length and SOL Thickness**

The well-known 2-point model illustrates the potential of extended connection length, $L$, in respect of separating the temperature at the target, $T_t$, and upstream, $T_u$, through the relation

$$T_u^{7/2} = T_t^{7/2} + \frac{7}{2} \frac{q_\| L}{\kappa_{0e}} \qquad (1)$$

where $q_\|$ is the density of power parallel to the magnetic field, and $\kappa_{0e}$ is the electron parallel conductivity coefficient [9]. One of the appealing features of the Super-X type of geometry is the ability to increase the connection length along the field line from the mid-plane to target, $L$, through reducing poloidal field, as opposed to simply increasing the distance in the poloidal plane. The theoretical potential of increased connection length is supported by detailed SOLPS simulations [5, 10-12].

Poloidal flux conservation between mid-plane (upstream) and target, where the poloidal fields are essentially vertical and radial respectively (see the right in Figure 1) leads to the relationship between the radial field in the divertor region, $B_{Rd}$, and the vertical field at the mid-plane, $B_{Zm}$ of

$$B_{Rd} = \frac{\delta R_m}{\Delta Z_d} \cdot \frac{R_m}{R_d} \cdot B_{Zm} \qquad (2)$$

where the radial distance across the region of SOL at the mid-plane, $\delta R_m$, represents the same flux increment as the height of expanded flux in the divertor, $\Delta Z_d$, with the average radii at the mid-plane and in the expanded region of the divertor being $R_m$ and $R_d$ respectively.

From this, the amount of the connection length, $L_d$, that is due to the field line traversing the divertor chamber is approximately given by

$$L_d \approx \frac{\Delta Z_d \Delta R_d}{\delta R_m} \frac{B_{\phi m}}{B_{Zm}} \qquad (3)$$

where $B_{\phi m}$ is the toroidal field at the mid-plane ($B_\phi \propto \frac{1}{R}$). Hence the useful connection length that can be achieved within a given divertor volume is inversely related to the thickness of the SOL at the mid-plane that is to be mapped magnetically through the divertor chamber to the target.

In MAST-Upgrade, the available volume of the divertor is represented by a chamber of height ~0.4m and radial extend ~0.4m whilst the ratio of the fields at the mid-plane for a high $\beta_p$ 1MA plasma is ~ 1.4. Hence from equation (3), the connection length addition from the divertor chamber is $L_d \approx \frac{0.22}{\delta R_m}$. Assuming that 1cm of the mid-plane SOL is carried through the divertor chamber then the maximum increment in connection length is available is ~22m. For a 2 cm thickness the increment is a more modest 11m. These numbers are in agreement with the equilibrium data shown in Figure 2.

**Divertor Closure**

The location of the nose is a balance between recycling from outside the divertor, and preventing the escape of neutrals from the divertor volume. This has been assessed with: heuristic analysis, coupled B2-EIRENE, SOLPS [13], EIRENE [14] calculations against a fixed plasma background (for fast assessment of different designs), and Monte-Carlo filament calculations [15]. All confirm that the design is a good balance – retaining neutrals in the divertor, without excessive recycling on the nose. Compression ratios (divertor /main chamber molecule density) between $10^2$ and $10^3$ are sought, as reported in C-MOD [16]. Hence the divertor structure is required to have an effective leakage area of order or less than 1 part in 1000 of the area of the gas-sealing surface, i.e. $< 10^{-2}$ m$^2$, in line with EIRENE and heuristic estimates, and in keeping with gas balance analysis reported by Maddison *et al* [17].

Figure 3a shows two different nose locations that were compared with SOLPS. In each case the flux of plasma leaving the mesh in the direction of the outer wall was simulated to be recycled from the divertor structures above the nose of the divertor, thereby providing a representation of the plasma that would be carried there along the field lines not captured in the SOLPS meshing. The effect of this is seen in Figure 3b, in which the mid-plane profiles of density from 3 pairs of SOLPS runs are compared. The 3 sets of diffusivities and separatrix densities were chosen to scan from narrow H-mode like profiles to L-mode type higher density plasmas. The recycling from the nose is seen in the outer parts of the profiles, but otherwise the plasma is little affected by the nose position. However with the nose closer to the plasma, the neutral density in the main chamber, at the wall, is lower by a factor of >2 (Table 1) showing that loss of neutrals from the divertor is more sensitive to nose position than recycling. Hence the final design has the closer-fitted nose, with tiles shaped to conform to typical flux surface geometry.

## Control of Divertor Configuration

Without high non-inductive current-drive fractions there will be a considerable variation of the solenoid current through the plasma pulse and this will affect the geometry of the Super-X type configurations. Figure 4 shows the difference in terms of the configuration and connection length at two representative solenoid currents. Zero solenoid current occurs early in the plasma current flat-top, whilst the -20kA case would be later in the pulse. Early in the pulse an X-point in the poloidal field lies close to the lower section of the divertor volume, and the flux surfaces seen from above could be described as concave. Later the flux surfaces are more convex, though the connection length is largely maintained. Such evolution drives the specification of: the divertor field power supplies, diagnostic sight lines, and shaping of plasma facing surfaces.

The operation requirements have been explored in terms of: plasma current, solenoid current, and a measure of the mid-plane SOL thickness that is connected along flux surfaces to the Super-X target tiles. This third variable is shown above to be inversely related to the connection length, but is a more directly accessible control variable in the code used to perform the analysis.

Example results are shown in Figure 5 in which the operating space for a 1MA high performance plasma is shown in terms of the solenoid current and magnetically-mapped SOL thickness. The solenoid current varies from positive to negative through the pulse, with a 1MA plasma current requiring the solenoid to have dropped to 20kA or less in order for sufficient flux to be transferred from the solenoid to the plasma. The full range of the solenoid is from 55kA down to -55kA, and the operating space demonstrates that the machine design supports operation over a wide range of solenoid swing and flux expansion of the SOL in Super-X geometries, especially as the measure of SOL thickness used here may be improved upon with changes to the magnetic configuration.

Operation with greatly expanded flux in the divertor region places demands on the accuracy with which the poloidal fields need to be controlled. This amounts to requiring the coil currents to be controlled within tens of amps, whilst the total currents are up to ~+/- 10kA. This requirement is expected to be achieved within the amplifiers themselves for the high frequency part, whilst the digital control system will need to enact this level of control for changes on the slower timescales of order 10ms and longer.

**Divertor Physics Exploration**

The MAST upgrade divertor will investigate the dependence of the SOL and edge plasma on connection length and neutral density around the main plasma. This can be expressed functionally as e.g. determining for the outer mid-plane SOL the density and temperature dependencies $n_e(r,L)$, $T_e(r,L)$ and $T_i(r,L)$, for the target the heat flux profile $q(s,L,P_{rad})$, and for the edge plasma $T_{e,ped}(P_{input},n_0)$ and $P_{L-H}(L)$, $f_{ELM}(P_{loss},L)$, with the connection length varying over a factor of 3 and the neutral density varying by over a factor of 10.

The aim is to provide well diagnosed experiments in the specific environment of MAST, and together with that, to develop theory and modelling that move forward both the understanding of the divertor physics processes, and the predictive capability for future machine designs, when combined with results from other machines using different wall materials and core and divertor plasma configurations.

**Conclusions**

The divertor design of the upgrade to MAST has been analysed in respect of key operational parameters. This drives the specification including: divertor field power supplies (both maximum current and regulation), divertor coils, divertor gas sealing, and nose location. The analysis provides insight into the issues that will need to be considered for including extended and expanded divertor configuration in DEMO-scale devices. Experiments in the upgraded MAST will assess the merits of such divertors.

## Acknowledgements

This work was part-funded by the RCUK Energy Programme under grant EP/I501045 and the European Communities under the contract of Association between EURATOM and CCFE. The views and opinions expressed herein do not necessarily reflect those of the European Commission.

Table 1 Parameters for SOLPS runs shown in figure 3, with 5MW of heating into a 1MA plasma. Transport values at separatrix, molecular densities at the main chamber wall (and divertor).

| Case | $n_e$ (m$^{-3}$) | $\chi_{i,e}$ (m$^2$s$^{-1}$) | D (m$^2$s$^{-1}$) | $n_{D2}$ (m$^{-3}$) close nose | $n_{D2}$ (m$^{-3}$) nose pulled back |
|---|---|---|---|---|---|
| H-mode | $1 \times 10^{19}$ | 1 | 0.1 | $1.1 \times 10^{17}$ ($1.8 \times 10^{19}$) | $3.0 \times 10^{17}$ ($1.3 \times 10^{19}$) |
| Mid | $1.5 \times 10^{19}$ | 2.5 | 1 | $1.8 \times 10^{17}$ ($7.3 \times 10^{19}$) | $4.3 \times 10^{17}$ ($5.5 \times 10^{19}$) |
| L-mode | $3 \times 10^{19}$ | 5 | 2 | $13 \times 10^{17}$ ($35 \times 10^{19}$) | $36 \times 10^{17}$ ($29 \times 10^{19}$) |

Figure 1. Representative up/down symmetric magnetic equilibria for the upgrade of MAST. Annotation referred to in the text.

Figure 2. Representative magnetic equilibria with full toroidal field (0.8T vacuum at R=0.8m) showing conventional (black), expanded strike point (red) and Super-X like long connection length (blue) divertor geometries.

Figure 3. a) Equilibria with close fitting nose (red) and nose pushed back (blue). b) Mid-plane density profiles for comparison cases. See table 1 for details of SOL parameters.

Figure 4. Like Figure 1 but with only 1cm of SOL mapped into the divertor and the data in the righthand plot are from the 5mm flux surface. Black outline is with no current in the solenoid, red outline is with -20kA in the solenoid.

Figure 5. Example operating space (inside red outline) in plasma current and magnetically-mapped mid-plane SOL thickness for a 1MA plasma with Super-X divertor configuration. The lines in the figure represent the limits of the various power supplies. The two diamonds correspond to the two configurations shown in Figure 4.

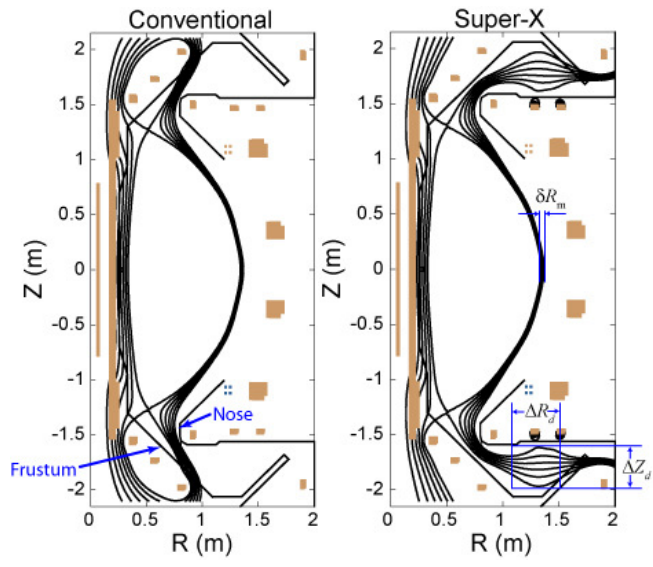

Figure 1.

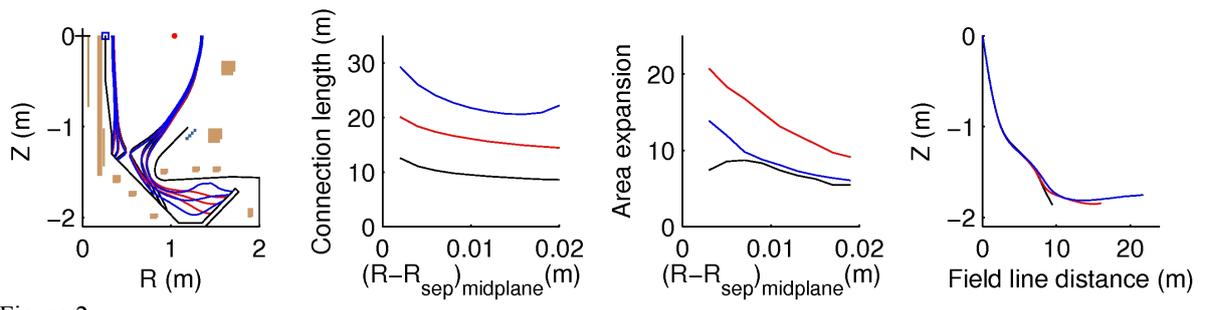

Figure 2.

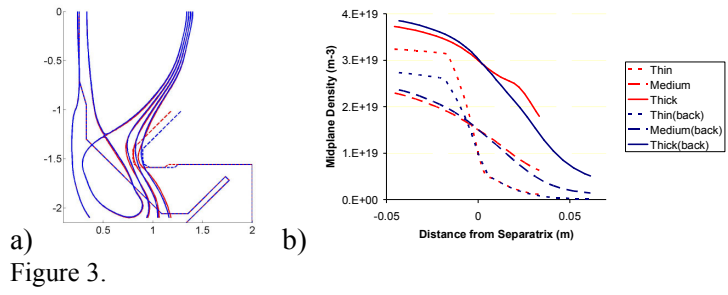
a) b)
Figure 3.

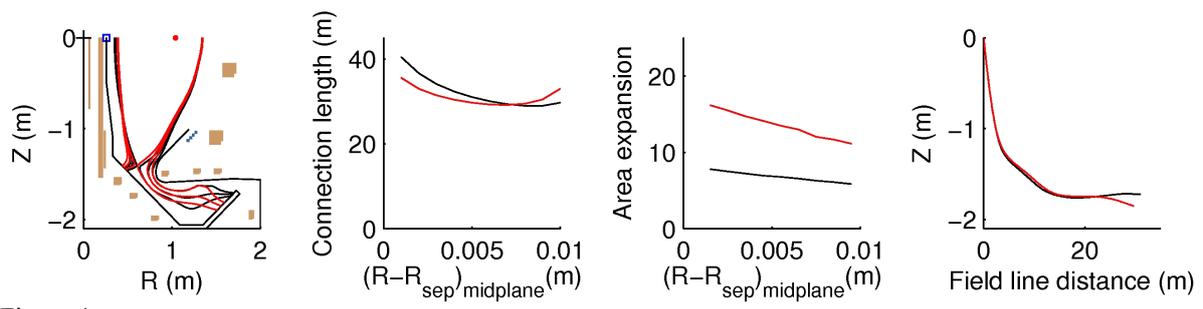

Figure 4.

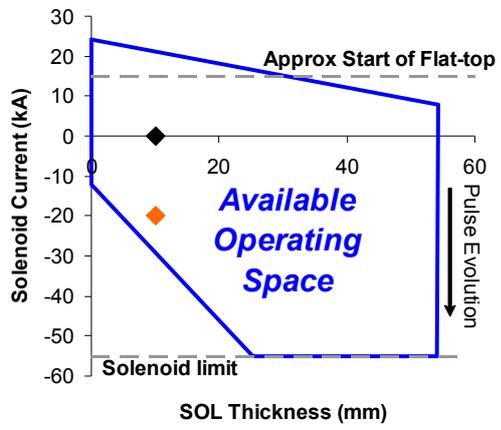

Figure 5.